\documentclass[twocolumn,noshowpacs,floatfix,groupedaddress]{revtex4}
\usepackage{graphicx}
\usepackage{epstopdf}
\usepackage{amssymb}
\usepackage{dcolumn}
\usepackage{bm}
\usepackage{subfigure}
\usepackage{color}
\usepackage{textcomp}
\usepackage{amsmath}
\usepackage{bbold}
\usepackage{multirow}

\newcommand{\COMMENTED}[1]{}
\newcommand{\REMARKS}[1]
{
{ \color{red}{\textbf{ {[#1]} }} }
}

\DeclareMathOperator{\Tr}{Tr}

\begin{document}

\title{Quantum Monte Carlo Simulation with Hartree-Fock-Bogoliubov Wave Function}

\title{Many-body computations by propagating Hartree-Fock-Bogoliubov wave functions}

\title{Many-body computations by random walks in Hartree-Fock-Bogoliubov space}

\title{Many-body computations by stochastic sampling in Hartree-Fock-Bogoliubov space}

\author{Hao Shi}
\author{Shiwei Zhang}

\affiliation{Department of Physics,
             The College of William and Mary,
             Williamsburg, Virginia 23187}

\begin{abstract}
% The Hartree-Fock-Bogoliubov (HFB) theory is a mean-field method which is widely applied in nuclear systems.
% We talk about Quantum Monte Carlo simulations using HFB wave function, show that both product state and Thouless state
% HFB wave function can be propagated by a generalized one-body operator. Illustrative examples on Kitaev is presented.
% The method study the Hamiltonian involving pairing fields or without U(1) symmetry, and pave a way to exactly solve many-body pairing problems. 

% The Hartree-Fock-Bogoliubov (HFB) theory is a mean-field method which is widely applied in nuclear systems. 
% By sampling HFB wave functions, we propose a many-body Quantum Monte Carlo algorithm. We show both product state and Thouless state can be propagated by a 
% generalized one-body operator, as well as been stabilized. Exact results after projection in Kitaev model is presented. Our method is numerical exact and size 
% scalable in strongly correlated systems, it pave the way to solve many-body problems involving pairing fields.

We describe the computational ingredients for an approach to treat interacting fermion systems in the presence of 
pairing fields, based on path-integrals in the space of  
Hartree-Fock-Bogoliubov (HFB) wave functions. The path-integrals can be evaluated by Monte Carlo, 
via random walks of HFB wave functions whose orbitals evolve stochastically.
%quantum Monte Carlo (QMC) algorithm by sampling Hartree-Fock-Bogoliubov (HFB) wave functions, which 
The approach combines the advantage of HFB theory in paired fermion systems 
and many-body quantum Monte Carlo (QMC) 
techniques. The properties of HFB states, written in the form of either product states or Thouless states, are
discussed. The states preserve forms when 
%shown to be 
propagated by generalized one-body operators. They %The states
can be stabilized for numerical iteration.
Overlaps and 
one-body Green's functions  between two such states can be computed. 
A constrained-path or phaseless approximation can be applied to the random walks of the HFB states
if a sign problem or phase problem is present.
The method is illustrated with an exact numerical projection in the Kitaev model, and in the Hubbard model 
with attractive interaction under an external pairing field. 

%Exact results after projection in Kitaev model is presented. Our method is size scalable in strongly correlated systems and pave the way to solve many-body problems involving pairing fields.

\end{abstract}

\maketitle

\section{Introduction}
\label{sec:intro}

%\textbullet Widely used HFB theory

For many-fermion systems with paring, 
the Hartree-Fock-Bogoliubov (HFB) approach \cite{HFB-book-1980} has been a key theoretical and computational tool. 
The approach has seen successful applications in the study of ground and certain excited states in nuclear systems, as well as in condensed matter 
%It is becoming more popular for interacting many-body systems in other fields, such as condensed-matter 
physics and quantum chemistry.
The method captures pairing and deformation correlations, and often provides a good symmetry-breaking picture for  weakly interacting systems. Symmetry
can also be restored by projection \cite{Gus-HFB-2011,PhysRevLett.108.042505} on a HFB vacuum, which further improves the quality of the approximation.

%\textbullet Lack many body algorithms1

For strongly interacting many-body systems, the HFB approach is not as effective, because of its underlying mean-field approximation.
%does not work well since it is eventually a mean-field method 
There have been attempts
to incorporate many-particle effects \cite{Imada-2008}. However a correlated HFB approach is still lacking which 
is size-consistent and scales 
in low polynomial computational cost with system size.

%\textbullet QMC is a powerful, however not direct target pairing

Quantum Monte Carlo (QMC) methods, which in general are scalable with system size, are among
the most powerful numerical approaches for interacting many-fermion systems. 
%is size-scalable in computing properties of strongly-interacting systems. It has been the a great sucess for a variety of boson systems
%and some fermion systems, and usually present benchmark for other methods \cite{Hubbard}. 
They have been applied in a variety of systems, including in systems where pairing is important. 
In such cases, HFB or related forms have been adopted as trial wave functions, for example, in 
diffusion Monte Carlo  (DMC) \cite{PhysRevLett.96.130201, Sandro-BCS-DMC} and  auxiliary-field QMC (AFQMC) \cite{PhysRevA.84.061602} calculations.
The HFB is used to provide a better approximate trial wave function with which to  guide the random walks 
by importance sampling, and to constrain the random walks 
if a sign problem %in order to control the sign problem if the latter
is present.  The random walks in these calculations do not sample HFB states, however; instead they 
take place in more ``conventional'' basis space, namely fermion position space in DMC or
Slater determinant space in AFQMC. 

The motivation for this paper is to formulate an approach which combines HFB with stochastic sampling.
From the standpoint of HFB theory, such an approach would provide a way to incorporate effects beyond
mean-field, by expressing the many-body solution as a linear combination of HFB states. From the 
standpoint of QMC, such an approach would allow 
the random walks to take place in the manifold of HFB states, 
which may provide a more compact representation of the interacting many-body wave function, especially 
in strongly paired fermion systems. To conduct the sampling in a space that
represents the many-body wave function or partition function more compactly generally  improves Monte Carlo efficiency (i.e., reduces statistical fluctuation
for fixed computational cost). Moreover it may  reduce the severity of the fermion sign/phase problem. 
%$\Psi_0\propto \sum_\psi w_\psi |\psi\rangle$, 
 %
 
 A further reason for developing such an approach is that present QMC methods generally 
 are not set up for many-body Hamiltonians which contain explicit
 pairing fields. 
 Such Hamiltonians can arise % in a variety of situations, 
 %are common in models for studying superconductors, 
 %for example, 
 in models for studying superconductors.
They can also arise from 
standard electronic Hamiltonians when a symmetry-breaking 
pairing field is applied to detect 
superconducting correlations.
Alternatively, if a pairing form of the Hubbard-Stratonovich (HS) transformation is applied 
to a standard two-body interaction, a bilinear Hamiltonian or action with pairing field will appear.
Moreover, when an electronic Hamiltonian is treated by an embedding framework \cite{Boxiao-2016},
the system is mapped into an impurity whose effective Hamiltonian is coupled to a bath and 
can break $U(1)$ symmetry. The impurity solver in that case would need to handle pairing fields.

%To solve these problems with QMC requires a method that can 
%handle many-fermion Hamiltonians without $U$(1) symmetry. 
 
\COMMENTED{
embedding 
  it may be required when the system has pairing fields or
in embedding framework.\cite{?}

It may be advantageous to have 
for example, in diffusion Monte Carlo calculations \cite{Schmidt, Sorella} which sample the wave functions by random walks in fermion 
positions, or in auxiliary-field QMC \cite{FG-3D} which samples the wave function by random walks in 
Slater determinant space. 

. However,
there may be situations where is is advantageous to take HFB as random samples, e.g., in strongly paired states. Further, it may be required when the system has pairing fields or
in embedding framework.\cite{?} 
}

%\textbullet Our proposal

In this paper we describe a QMC method for handling many-fermion Hamiltonians without $U$(1) symmetry. 
The method evaluates the path integral in auxiliary-field space to produce a ground-state wave function 
(or finite-temperature partition function) by sampling HFB states. 
%The method 
It is a generalization of the AFQMC method from the space of Slater determinants (Hartree-Fock
states) to that of HFB states. Below we formulate the QMC approach in this framework, and then outline
all the ingredients for implementing a computational algorithm. 
We illustrate the method with two examples. The first is a solution of the Kitaev model 
by imaginary-time projection. This is a non-interacting problem whose ground state is available exactly,
and serves as an excellent toy problem for illustrating the key elements of the method. The 
second example is the attractive Hubbard model. We study the pairing order in this model
by applying an explicit pairing field that breaks particle number symmetry.

\COMMENTED{
In this paper, we propose a QMC algorithm by sampling HFB wave function.  The method deals with paring operator directly, thus solve the Hamiltonian without U(1) symmetry. 
It captures physics in strongly-correlated pairing systems.
}

The remainder of this paper is organized as follows. In Sec.~\ref{sec:QMC} we summarize the QMC 
formalism by highlighting all the ingredients necessary for an efficient sampling of the HFB space. 
In Sec.~\ref{sec:HFB} we give a brief introduction of the standard HFB approach to facilitate the ensuing 
discussion. In Sec.~\ref{sec:method} we present our method. The random walkers can take either
of two forms of HFB state, a product state or a so-called Thouless state, and they are discussed separately.
Then in Sec.~\ref{sec:result} we present our illustrative results on the Kitaev model and on the 
attractive Hubbard model. Finally in Sec.~\ref{sec:sum} we conclude with a brief discussion and 
summary.

\section{QMC FORMALISM}
\label{sec:QMC}

In this section we briefly outline the key steps in the 
%We briefly talk about standard 
ground-state AFQMC method, to facilitate the discussion of propagating an HFB wave function. 
We will use the open-ended branching random walk approach \cite{AFQMC-lecture-notes-2013}; however, the 
alternative of 
Metropolis sampling of a fixed (imaginary-)length path integral  \cite{assaad-lec-note} shares the same algorithmic ingredients
in the context of formulating an approach with HFB wave functions.
%all key ingredients are the same as those in Metropolis algorithms  \cite{?}. 
%To learn more details about the algorithms, please refer to \cite{?}.
Additional details of the AFQMC methods can be found in Refs.~\cite{AFQMC-lecture-notes-2013, assaad-lec-note}.

Imaginary-time
projection is a common way to solve the ground state of many-body problems.
%one of the most popular methods to solve many-body problems.  
The ground state 
wave function $|\Psi_0 \rangle$ of Hamiltonian $\hat{H}$ is projected out by
\begin{equation}
 |\Psi_0 \rangle \propto \lim_{\tau\rightarrow \infty} \exp(-\tau \hat{H}) |\psi_T \rangle\,,
\end{equation}
where the initial state $| \Psi_T \rangle$, which we will take to be the same as the trial wave function, is not orthogonal with $| \Psi_0 \rangle$. The long imaginary time $\tau$ is divided into smaller steps
(each referred as a time slice): $\tau = L \Delta \tau$, and 
\begin{equation}
 \label{eq:Trotter}
 \exp(-\tau \hat{H}) = \prod_{l=1}^L \exp(-\Delta\tau \hat{H})\,.
\end{equation}
By using the Trotter-Suzuki breakup and Hubbard-Stratonovich (HS) transformation, the projection operator can be expressed in an integral form
\begin{equation}
 \label{eq:Bx}
 \exp(-\Delta \tau \hat{H}) \doteq\int  p(x) \exp[\hat{O}(x)] dx\,,
\end{equation}
where the auxiliary-field $x$ is a vector whose dimensionality is typically 
proportional to the size of the basis $N$,
$p(x)$ is a probability density function, and 
$\hat{O}(x)$ is a one-body operator containing terms of order $\Delta\tau$ and  $\sqrt{\Delta\tau}$. The residual Trotter errors in Eq.~(\ref{eq:Bx}) are higher order in $\Delta\tau$, which are removed in practice by 
choosing sufficiently small time-steps and extrapolation with separate calculations using different values 
of $\Delta\tau$.
There are different HS fields which couple with spin, charge, or pairing operators. These fields lead
to different forms of $\hat{O}(x)$, which can be Hartree, Hartree-Fock, or pairing form. The form of the HS 
affects the efficiency of the QMC algorithm, as well as the systematic accuracy if a constraint is applied
to control the sign or phase problem \cite{shiweiPRL1995, PhysRevLett.90.136401}. We will not be concerned with the details here as they have minimal effect 
on the formalism below.

%\REMARKS{Can not define the general operator here, since we do not define creation and destroy operator, move to method section.}

Formally the many-body 
ground state wave function can be expressed as a high-dimensional integral:
\begin{equation}
|\Psi_0 \rangle \propto \idotsint \prod_{l=1}^L d{x_l} p(x_l)\,|\psi_{\pmb X} \rangle\,,
\end{equation}
where $l$ denotes a time slice as in Eq.~(\ref{eq:Trotter}), and
 %with
\begin{equation}
 |\psi_{\pmb X} \rangle =\prod_{l=1}^L \exp[\hat{O}(x_l)] |\psi_T \rangle\,.
\end{equation}
 The shorthand $\pmb {X}$ denotes the collection of the HS fields along the imaginary-time path, 
 $\{x_1, x_2, \ldots , x_L\}$.
It can be sampled by QMC, either via branching random walks or the Metropolis algorithm,
% technique to a assembly 
to give, formally:
\begin{equation}
\label{eq:Psi0-MC}
|\Psi_0 \rangle \propto \sum_{\pmb X} W_{\pmb X}  |\psi_{\pmb X} \rangle,
\end{equation}
where $W_{\pmb X}$ is a Monte Carlo weight for  $\pmb {X}$ (which can depend on the importance
sampling transformation \cite{AFQMC-lecture-notes-2013}).

The above assumes that 
\begin{equation}
 \label{eq:PropStep}
 \exp[\hat{O}(x)] |\psi \rangle \rightarrow |\psi' \rangle\,,
\end{equation}
i.e. the action by the propagator of Eq.~(\ref{eq:Bx}) on a state leads to a new state of the same 
form. 
%One important requirement is that $|\psi_{\pmb X}\rangle$ must has the same form as $|\psi_T \rangle$, i.e., 
If $|\psi\rangle$ is a Slater determinant, a coordinate space state,
or a matrix product state \cite{PhysRevB.90.045104}, then $|\psi'\rangle$ has the same respective form.
For example, in AFQMC the states are single Slater determinants and $\exp[\hat{O}(x)] $ is 
a one-body propagator, while in DMC the states are a collection of particle positions and the 
propagator is a translation operator.
Below we will assume that 
$|\psi_T \rangle$ is an HFB wave function (or a linear combination of HFB states), and show that
we can generalize Eq.~(\ref{eq:PropStep}) to HFB states and turn the propagation into 
a random walk in HFB space.

With Eq.~(\ref{eq:Psi0-MC}) we can make measurements of the ground state energy by:
%The ground state energy can be measured by
\begin{equation}
 E_0= \frac{\sum\limits_{\pmb X} W_{\pmb X}  \langle \psi_T |\hat{H} |\psi_{\pmb X} \rangle }{ \sum\limits_{\pmb X} W_{\pmb X}  \langle \psi_T |\psi_{\pmb X} \rangle}\,.
\end{equation}
Other observables (that do not commute with the Hamiltonian) and correlation functions can be measured by back-propagation \cite{PhysRevB.55.7464, PhysRevE.70.056702}. 
In the Metropolis approach where the entire path is kept, measurement 
can be carried out in the middle portion of the path. (This may lead to an infinite variance problem 
which can be controlled \cite{PhysRevE.93.033303}.)

%There are a few ingredients to form a sucessful 
We can list all the key ingredients needed in the QMC algorithm:
\begin{enumerate}
 \item  The random walker $ |\psi\rangle$, when propagated by the operator $ \exp(\hat{O})$ in Eq.~(\ref{eq:Bx}),
 evolves into another state, $ |\psi'\rangle$, of the same form, as in Eq.~(\ref{eq:PropStep}).
 \COMMENTED{
 the Eq.~(\ref{})
 Apply an exponential one-body operator to a wave function $|\psi^\prime \rangle= \exp(\hat{O}) |\psi\rangle$, where $|\psi^\prime \rangle$ and $|\psi \rangle$ has the same form. }
 
 \item  The overlap of two ``walker'' wave functions, 
 $\langle \psi^\prime|\psi\rangle$, needs to be calculated (in low polynomial complexity).
 
 \item  The Green's function given by a quadratic operator  $\hat{C}$ needs to be computed,
  $\langle \psi^\prime|\hat{C}|\psi\rangle /\langle \psi^\prime|\psi\rangle$, again with low polynomial complexity. 
  In addition, correlation functions 
  (quartic operators) need to be computed from these (as in Wick's theorem with Slater determinants). 

\item The walker wave function need to be stable (or stabilized) numerically during long imaginary-time 
propagation.

\end{enumerate}

With these ingredients, force bias can be computed \cite{AFQMC-lecture-notes-2013, PhysRevA.92.033603}
to allow importance sampling to achieve better efficiency. Symmetry properties can be imposed \cite{PhysRevB.88.125132, PhysRevB.89.125129, PhysRevLett.111.012502}.
A constrained-path \cite{shiweiPRL1995} or phaseless \cite{PhysRevLett.90.136401}
approximation can be introduced to control the sign problem. 
A full AFQMC-like computation can then be carried out, following either the
Metropolis path-integral procedure (including force bias), or with  open-ended random walks and a
constraint if there is a sign or phase problem.

%\REMARKS{after the structure is fixed, mention symmetry, infinite variance in QMC, mention new algorithm can be applied for fermi gas systems.}

%\section{HFB Formalism}
\section{HFB Basics}

\label{sec:HFB}

Let us first define a set of $N$ single particle creation operators, $c^{\dagger}=\begin{pmatrix}c_1^{\dagger} &c_2^{\dagger}&\dots& c_N^{\dagger}\end{pmatrix}$,  and annihilation operators, $c=\begin{pmatrix}c_1 &c_2 &\dots& c_N\end{pmatrix}$,
which satisfy fermion commutation relations. Quasi-particle bases $\beta^{\dagger}$ and $\beta$, with the same form as $c^{\dagger}$ and $c$,  can be set through a unitary Bogoliubov transformation, 
\begin{equation}
\label{eq:quasi-part-op-def}
    \begin{pmatrix}
        \beta^{\dagger} & \beta
     \end{pmatrix}
     = 
    \begin{pmatrix}
     c^{\dagger}  & c
    \end{pmatrix}
     \begin{pmatrix}
     \mathbb{U}  & \mathbb{V}^* \\
     \mathbb{V}  & \mathbb{U}^*
    \end{pmatrix},
\end{equation}
Here $\mathbb{U}$ and $\mathbb{V}$ are $N \times N$ matrices.
For example,  $N=2\,N_{\rm basis}$ for spin-$1/2$ fermions in a basis of size $N_{\rm basis}$.

The vacuum of quasi particles  is an HFB wave function. It can be written in the form of a \emph{product state}, with annihilation operators $\beta$ applied to the true vacuum,
\begin{equation}
\label{eq:product-state}
|\psi_p \rangle =\prod_i^N \beta_i |0 \rangle\,,
\end{equation}
where
the quasi-particle operator  $\beta_i$ is 
the $(N+i)$-th element of the vector on the left-hand side in Eq.~(\ref{eq:quasi-part-op-def}).
In the case of a fully paired state when $\mathbb{U}$ is invertible,
an HFB state can alternatively be expressed in the form of a \emph{Thouless state}:
\begin{equation}
\label{eq:Thouless-state}
% |\psi_t \rangle= \exp(\frac{1}{2} c^{\dagger} \mathbb{Z} c^{\dagger} ) |0 \rangle
|\psi_t \rangle= \exp(\frac{1}{2} c^{\dagger} \mathbb{Z} (c^{\dagger})^T ) |0 \rangle\,,
\end{equation}
where $\mathbb{Z}=(\mathbb{V} \mathbb{U}^{-1})^*$, 
and the superscript ``$T$'' indicates ``transpose''.

When both exist, the two forms are connected by a simple relation $|\psi_p \rangle = \mathrm{pf}(\mathbb{U}^\dagger \mathbb{V}^*)|\psi_t \rangle $, where `pf' denotes Pfaffian (see below).
%It is okay to choose either form in QMC simulations, since they are connected by a simple relation $|\psi_p \rangle = \mathrm{pf}(\mathbb{U}^\dagger \mathbb{V}^*)|\psi_t \rangle $. 
%Product state reduces to a Slater determinant state when $\mathbb{U}=\mathbb{0}$. 
%(A singular problem arises due to full occupation ($\det(\mathbb{U})=0$), which can be easily fixed \cite{Gao2014360}.) 
%(A singularity problem arises due to full occupation, $\det(\mathbb{U})= 0$.
% which can be easily fixed 
%Ref.~\cite{Gao2014360} discusses a numerical trick which may be helpful .) 
%
In Sec.~\ref{sec:method} we discuss the QMC formalisms based on each of these two forms
as random walkers. 

\COMMENTED{
We first use product state to illustrate a typical QMC procedure. When $\mathbb{U}$ is alway invertible, Thouless state wave function is faster since it has smaller matrix size and automatically stabilization procedure. We talk about details in section \ref{sec:method}. 
}

\section{Method}
\label{sec:method}

In this section,
we show how the four ingredients for a QMC simulation listed in Sec.~\ref{sec:QMC} can be realized 
with HFB states. We first discuss product states in Sec.~\ref{subsec:product}, which are formally a more direct generalization of Slater 
determinants in AFQMC. This is followed in the next section by the details for Thouless 
states. When $\mathbb{U}$ is invertible, Thouless states are faster  than product states, since they have smaller matrix size and an automatic stabilization procedure, as illustrated in Sec.~\ref{subsec:Thouless}.
\COMMENTED{
four technique elements talked in section \ref{sec:QMC} for QMC simulation with HFB wave function. 
Since we need the formalism of overlap for propagating HFB wave function, we first talk about overlap, followed by Green's function, then propagating HFB wave function and stabilization.}
Some mathematical details are left to the Appendix, in order to not impede the flow of the discussion.
%which might break the flow of the presentation is left in Appendix.

We write, without loss of generality, the one-body operator  $\hat{O}$ in Eq.~(\ref{eq:Bx}) that results after HS transformation
of the interacting Hamiltonian in the following form:
%General Hermition operator $\hat{O}$ is used to propagate HFB wave function,
\begin{equation}
\label{eq:general-O-def}
 \hat{O} = \sum_{ij}^{N} t_{ij} c_i^\dagger c_j + \sum_{i>j}^{N} \Delta_{ij} c_i c_j + \sum_{i>j}^{N} \widetilde{\Delta}_{ij} c_i^\dagger c_j^\dagger, 
 %\hat{O} = \sum_{ij}^{N} t_{ij} c_i^\dagger c_j + \sum_{i>j}^{N} \Delta_{ij} c_i c_j + \sum_{i>j}^{N} \Delta_{ji}^* c_i^\dagger c_j^\dagger,
\end{equation}
With $\Delta^T=-\Delta$ and $\widetilde{\Delta}^T=-\widetilde{\Delta}$. %Note that the operator is a special form of the one described in Appendix \ref{app:addformula}.

\COMMENTED{
In section \ref{subsec:product},  we talk about details for product states. 
In section \ref{subsec:Thouless}, we talk about details for Thouless states.
Finally in section \ref{subsec:example}, illustrative results are discussed.
}

\subsection{Product state}
\label{subsec:product}
\textit{Overlap}: 
In a QMC simulation, we need to calculate the overlap of two HFB wave functions. 
With importance sampling, typically only the ratio of overlaps are needed, for example,
%These overlaps are typically needed in the form of a ratio, for example,
$\langle \psi_T | \exp(\hat{O}) |\psi_p\rangle/\langle \psi_T |\psi_p\rangle$, 
where $|\psi_T\rangle$ is the trial wave function.
\COMMENTED{
The sign/phase of the overlap must be correctly determined (at least the relative sign/phase in the ratio) for 
imposition of constrains \cite{?}. 
}
Onishi's Theorem provides a simple way to calculate
\begin{equation}
\label{eq:Onishi}
\langle \psi_p| \psi_p^\prime \rangle^2= \det(\mathbb{U^{\prime\dagger} U} + \mathbb{V^{\prime\dagger} V}) \det(\mathbb{V^{\prime\dagger} V} )\,, 
\end{equation}
where $\mathbb{U}$ and $\mathbb{V}$ are the components of the unitary transformation matrix of  $|\psi_p\rangle$ as defined earlier, and
 $\mathbb{U}^{\prime}$ and $\mathbb{V}^{\prime}$ are those
  %in the unitary transformation matrix 
  for $|\psi_p^{\prime}\rangle$.
  This formula can be used to evaluate the normalization of a product state, for example. 
However, Eq.~(\ref{eq:Onishi}) 
 neglects the sign in the overlap.
 The sign/phase of the overlap is important (at least the relative sign/phase in the ratio above) in order to  
impose the constraint to control  the sign or phase problem \cite{AFQMC-lecture-notes-2013}. 
%  we can only use this formula when $| \psi_p^\prime \rangle =| \psi_p\rangle$.  L. M. 
Robledo worked out the following form \cite{PhysRevC.79.021302} which regains the sign of the overlap:
\begin{equation}
\label{eq:overlap-prod-pf}
 \langle \psi_p| \psi_p^\prime \rangle=  (-1) ^{N(N-1)/2} \mathrm{pf} 
      \begin{pmatrix}
     \mathbb{V}^T \mathbb{U}                 &   \mathbb{V}^T\mathbb{V}^{\prime*} \\
     -\mathbb{V}^{\prime \dagger}\mathbb{V}  &   \mathbb{U}^{\prime\dagger} \mathbb{V}^{\prime *}
    \end{pmatrix}\,,
\end{equation}
where the Pfaffian % (pf) 
can be computed (see, e.g., library by Bertsch \cite{GonzálezBallestero20112213}). Note that,
when $\mathbb{U}=0$,
 Eq.~(\ref{eq:overlap-prod-pf}) will reduce to the formula of Slater determinants, $\det(\mathbb{V^{\prime\dagger} V})$, as expected.

\textit{Green's Function}:
Physical properties are measured through Green's functions in AFQMC.
% it has been worked out in the past\cite{?}. 
Similar generalization can be made from Slater determinants to HFB product states.
Let us set $\mathbb{Q}=( \mathbb{U^{\prime\dagger} U} + \mathbb{V^{\prime\dagger} V} )^T$.
The three types of Green's functions  are then given by
\begin{eqnarray}
 \rho_{ij}=\frac{\langle \psi_p |c_i^{\dagger} c_j| \psi_p^\prime \rangle}{\langle \psi_p | \psi_p^\prime \rangle} &=& (\mathbb{V}^{\prime*} \mathbb{Q}^{-1} \mathbb{V}^T )_{ji}~,  \nonumber\\
 \kappa_{ij}=\frac{\langle \psi_p |c_i c_j| \psi_p^\prime \rangle}{\langle \psi_p | \psi_p^\prime \rangle} &=& (\mathbb{V}^{\prime*} \mathbb{Q}^{-1} \mathbb{U}^T )_{ji}~,  \nonumber\\
 \overline{\kappa}_{ij}=\frac{\langle \psi_p |c_i^{\dagger} c_j^{\dagger}| \psi_p^\prime \rangle}{\langle \psi_p | \psi_p^\prime \rangle} &=& -(\mathbb{U}^{\prime*} \mathbb{Q}^{-1} \mathbb{V}^T )_{ij}~.
\end{eqnarray}
Note that, when $\mathbb{U}=\mathbb{U}'=\mathbb{0}$, the first line reduces to the Slater determinant result, while the last two lines vanish, as expected.

A generalized Wick's theorem \cite{ONISHI1966367, Balian1969} holds, which 
allows expectation values of two-body operators and correlation functions to be calculated. For example,
\begin{equation}
 \frac{\langle \psi_p |c_i^{\dagger} c_j^{\dagger} c_k c_l| \psi_p^\prime \rangle}{\langle \psi_p | \psi_p^\prime \rangle} 
 = \rho_{il} \rho_{jk} -\rho_{ik} \rho_{jl} + \overline{\kappa}_{ij} \kappa_{kl}.
\end{equation}

\textit{Propagation}:
We need to apply the exponential of a general one-body operator $\hat{O}$ to a product HFB wavefunction.
It can be shown (see Appendix~\ref{app:addformula})
%From the Appendix \ref{app:lineartrans}, 
that $\exp(\hat{O})$ can be ``exchanged'' %permute 
 with a quasi-particle operator $\beta_i$ in the following manner 
\begin{equation}
\exp(\hat{O}) \beta_i  =\beta_i^\prime \exp(\hat{O}),
\end{equation}
i.e., by modifying $\beta_i$ to a new form  $\beta_i^\prime$ defined with
the matrix multiplication
\begin{equation}
\label{eq:update-prod-state}
 \beta^\prime =     
    \begin{pmatrix}
     c^{\dagger}  & c
    \end{pmatrix} 
    \exp
    \begin{pmatrix}
     t      & \widetilde{\Delta} \\
     \Delta & -t^T
    \end{pmatrix}
    \begin{pmatrix}
    \mathbb{V}^*\\
    \mathbb{U}^*
    \end{pmatrix}\,.
\end{equation}
Successive applications of the above 
%to the quasi-particle operators $\{\beta_i\}$ of the present product state $|\psi_p\rangle$ 
yields
%We can keep on permuting $\exp(\hat{O})$ in a product state,
\begin{equation}
\label{eqn:expotowf}
 \exp(\hat{O})\,\prod_i \beta_i |0\rangle = \prod_i \beta_i^\prime\,\exp(\hat{O}) |0\rangle\,.
\end{equation}
As shown in Eq.~(\ref{eqn:expand}) in the Appendix, $\exp(\hat{O}) |0\rangle$ on 
the right-hand side in Eq.~(\ref{eqn:expotowf}) can be written as
\begin{equation}
 \exp(\hat{O}) |0\rangle  \propto \exp[\frac{1}{2} c^\dagger \mathbb{Z}_0 (c^\dagger)^T] |0 \rangle,
\end{equation}
which gives quasiparticle states that are either paired or empty.  So the 
 right-hand side of Eq.~(\ref{eqn:expotowf}) 
is the vacuum of the new quasi-particle operator $\beta_i^\prime$, which 
is equivalent to $\prod_i \beta_i^\prime |0\rangle$ 
up to a constant factor: %except for normalization:  
%\REMARKS{Need to refine here.} %With these we get
\begin{equation}
 \exp(\hat{O}) \prod_i \beta_i |0\rangle = \alpha \prod_i \beta_i^\prime |0\rangle\,.
\end{equation}
The normalization $\alpha$ can be determined by
\begin{equation}
\label{eqn:alphaproduct}
\alpha =\frac{\langle \phi | \exp(\hat{O}) \prod_i \beta_i |0\rangle} {\langle \phi|   \prod_i \beta_i^\prime |0\rangle}\,,
\end{equation}
where $|\phi\rangle$ can be any state. For example, 
the calculation is straightforward when $|\phi\rangle$ is  chosen to be the true vacuum or an eigenstate of $\hat{O}$
(see Appendix \ref{app:addformula} for details). 
Note that $\alpha$ is always $1$ if there is no pairing operator, since $\exp(\hat{O}) |0\rangle =|0\rangle$.
This covers the case of the propagation of Slater determinants in standard AFQMC. 
It also includes, for example, the situation where a pairing trial wave function is used but 
to a Hamiltonian with no pairing field and a HS transformation that does not 
involve pairing decompositions. If pairing is between two spin components, we can choose the \textit{vacuum} to be the true vacuum of one
spin component, and ``fully occupied'' for the other spin component, which will reduce $\alpha$ to $1$.

\textit{Stabilization}:
A unitary Bogoliubov transformation imposes  fermion commutation relations to the quasi-particle operators,
which %guarantee that its 
%and 
ensures that the product form of the HFB wave function is well-defined. There are two stabilization conditions
\begin{equation}
\label{eqn:stable1}
\mathbb{U}^\dagger \mathbb{U}+ \mathbb{V}^\dagger \mathbb{V} = \mathbb{1}
\end{equation}
and
\begin{equation}
\label{eqn:stable2}
 \mathbb{U}^T \mathbb{V}+ \mathbb{V}^T \mathbb{U} = \mathbb{0}.
\end{equation}
During the  iterative propagation, the  transformation  matrices $\mathbb{U}$ and $\mathbb{V}$  are updated 
%to $\mathbb{U^\prime}$ and $\mathbb{V^\prime}$ 
following Eq.~(\ref{eq:update-prod-state}):
%through 
\begin{equation}
\label{eq:newProductMatrix}
     \begin{pmatrix}
     \mathbb{V^{\prime *}} \\
     \mathbb{U^{\prime *}}
    \end{pmatrix} 
    =
    \exp
    \begin{pmatrix}
     t      & \widetilde{\Delta} \\
     \Delta & -t^T
    \end{pmatrix}
    \begin{pmatrix}
    \mathbb{V}^*\\
    \mathbb{U}^*
    \end{pmatrix}.
\end{equation}

It is easy to show that, if $\hat{O}$ is Hermition, and  $\mathbb{U}$ and $\mathbb{V}$  satisfy the second condition above, Eq.~(\ref{eqn:stable2}), then the new 
matrices $\mathbb{U}^{\prime}$ and  $\mathbb{V}^{\prime}$ will follow the same condition.
%mathematically. 
However, these conditions can be violated if $\hat{O}$ has a general form, or simply 
because of numerical instabilities
caused by finite precision. %  $\hat{O}$.
%The only way to break Eq. (\ref{eqn:stable2}) is numerical instability, which 
This can be restored by forcing skew-symmetry to 
\begin{equation}
\mathbb{B}\equiv \mathbb{U}^{\prime T} \mathbb{V}^\prime\,,
\end{equation}
after which we modify $\mathbb{U}^{\prime T}$ if $\mathbb{V}^\prime$ is invertible, or vice versa.

The first condition is similar to the situation with Slater determinants in AFQMC. Single particle states created by the quasi-particle operators %$\beta_i^\prime$ need to be 
must remain orthonormal to each other. The propagation can violate this condition and cause numerical 
instability. This can be stabilized by, for example,
the modified Gram-Schmidt (modGS) procedure,
\begin{equation}
\label{eqn:mgs}
     \begin{pmatrix}
    \mathbb{V}^{\prime*}\\
    \mathbb{U}^{\prime*}
    \end{pmatrix}
     =
     \begin{pmatrix}
     \widetilde{\mathbb{V}}^{\prime*}\\
     \widetilde{\mathbb{U}}^{\prime*}
    \end{pmatrix}
    \mathbb{R}\,,
\end{equation}
where $\mathbb{R}$ is an upper triangular matrix, and $\det(\mathbb{R})$ represents the overall normalization/weight of the HFB wave function which usually needs to be stored. 
Similar to the modGS stabilization in AFQMC, the off-diagonal part of $\mathbb{R}$ represents nonorthogonality in the original quasi-particle basis,
which does not affect the %always contribute zero to 
HFB wave function, and can thus be discarded.

It is worth noting that we should always force skew-symmetry of $\mathbb{B}$ before applying 
the modGS %Modified Gram-schmidt 
process. %Since 
This is because changes in $\mathbb{B}$ will affect orthonormality, while %Modified Gram-schmidt 
the modGS will not change the skew-symmetry of $\mathbb{B}$: 
\begin{equation}
\widetilde{\mathbb{B}} = \widetilde{\mathbb{U}}^{\prime T} \widetilde{\mathbb{V}}^\prime = \mathbb{R}^{\dagger-1} ( \mathbb{U}^{\prime T} \mathbb{V}^\prime ) \mathbb{R}^{*-1}\,,
\end{equation}
i.e., $\widetilde{\mathbb{B}}$ has the same skew symmetry as $\mathbb{B}$.

\subsection{Thouless state}
\label{subsec:Thouless}

When a fully paired state is involved which allows the use of a Thouless form, similar formulas can be 
written down.

\textit{Overlap}:
The overlap of two Thouless states is \cite{PhysRevC.79.021302}
\begin{equation}
 \langle \psi_t| \psi_t^\prime \rangle=  (-1) ^{N(N+1)/2} \mathrm{pf} 
      \begin{pmatrix}
      \mathbb{Z^\prime}     &   -\mathbb{1} \\
     \mathbb{1}             &   -\mathbb{Z}^*
    \end{pmatrix}.
\end{equation}

\textit{Green's Function}:
With the same definition as in Sec.~\ref{subsec:product}, the Green's functions should be the same in 
the Thouless form as in product state form. %The formula 
They can be written more 
compactly for Thouless states: % is more elegant.
\begin{equation}
      \begin{pmatrix}
      \overline{\kappa}     &   \rho \\
      -\rho^T             &   \kappa
    \end{pmatrix}
    =
    \begin{pmatrix}
      \mathbb{0}     &      \mathbb{1} \\
      -\mathbb{1}             &   \mathbb{0}
    \end{pmatrix}
    -
    \begin{pmatrix}
      \mathbb{Z^\prime}     &   -\mathbb{1} \\
     \mathbb{1}             &   -\mathbb{Z}^*
    \end{pmatrix}^{-1}\,.
\end{equation}
The above can be shown using coherent states.  The ingredients are similar to those used in the evaluation of 
overlaps in Ref.~\cite{PhysRevC.79.021302}.

\textit{Propagation}:
%As defined in Appendix \ref{app:compress}, we have 
Let us denote the matrix representation of $\exp(\hat{O})$ by 
%\REMARKS{:necessary??}(see Appendix \ref{app:compress})
\begin{equation}
\exp(\mathbb{O})=
 \begin{pmatrix}
  \mathbb{K}   &  \mathbb{M}  \\
  \mathbb{L}   &  \mathbb{N}
 \end{pmatrix}\,.
\end{equation}
The application of $\exp(\hat{O})$  on the
Thouless state $|\psi_t\rangle$ gives
%is just another operator applied on true vacuum. Combining two operators as shown in Appendix \ref{app:compress},
%we get a new operator $\hat{O^\prime}$,
\begin{equation}
 \exp(\hat{O}) |\psi_t\rangle \propto\exp(\hat{O^\prime}) |0\rangle\,,
\end{equation}
after the one-body operator $\hat{O}$  is combined with the pairing operator from 
$|\psi_t\rangle$ (see Appendix~\ref{app:addformula}).
The corresponding matrix representation of the new operator $\hat{O^\prime}$ is given by
%for  $\exp(\hat{O^\prime})$ is
\begin{equation}
\exp(\mathbb{O^\prime})=
 \begin{pmatrix}
  \mathbb{K}   &  \mathbb{KZ+M}  \\
  \mathbb{L}   &  \mathbb{LZ+N}
 \end{pmatrix}\,.
\end{equation}
Using the expansion in Eq.~(\ref{eqn:expand}), we have
\begin{equation}
 \exp(\hat{O^\prime}) |0\rangle  \propto \exp(\frac{1}{2} c^\dagger \mathbb{Z}^\prime c^\dagger) |0 \rangle\,,
\end{equation}
with 
\begin{equation}
\mathbb{Z}^\prime= (\mathbb{KZ+M}) (\mathbb{LZ+N})^{-1}\,.
\end{equation}
The new Thouless wave function after propagation is
\COMMENTED{
\begin{equation}
 |\psi_t^\prime\rangle = \exp(\frac{1}{2} c^\dagger \mathbb{Z}^\prime c^\dagger)  |0\rangle\,.
\end{equation}
Up to now, we still miss a factor $\alpha$ in the propagation,
\begin{equation}
 \exp(\hat{O}) |\psi_t\rangle = \alpha  |\psi_t^\prime\rangle.
\end{equation}
}
\begin{equation}
|\psi_t^\prime\rangle \equiv \exp(\hat{O}) |\psi_t\rangle = \alpha \exp(\frac{1}{2} c^\dagger \mathbb{Z}^\prime c^\dagger)  |0\rangle\,.
\end{equation}    
The weight/normalization of the new  state can be determined by
\begin{equation}
\label{eqn:alphathouless}
 \alpha =\frac{\langle \phi| \exp(\hat{O}) |\psi_t\rangle }{  \langle \phi |\psi_t^\prime\rangle}\,,
\end{equation}
where we can choose, for example, 
%It is better to choose 
$|\phi\rangle =|0\rangle$, and use Eq.~(\ref{eqn:expand}) to expand $\exp(\hat{O})$ before calculating the overlap (see Appendix
\ref{app:addformula}).

\textit{Stabilization}:
As we stabilize the product state in Eq.~(\ref{eqn:mgs}), we have
\begin{equation}
 \mathbb{Z} = (\mathbb{V} \mathbb{U}^{-1})^* = (\widetilde{\mathbb{V}} \widetilde{\mathbb{U}}^{-1})^*,
\end{equation}
so that the 
matrix $\mathbb{R}$ cancels when the matrix $\mathbb{Z}$ is formed, and the Thouless state is 
unchanged.
This suggests that Thouless state is more stable during the propagation. 
Numerical instability can contaminate the HFB  
%The only thing might break the HFB 
wave function.
%is numerical instability, which can be easily fixed by forcing skew symmetry of $\mathbb{Z}$.
Skew symmetry of $\mathbb{Z}$ should be enforced to help maintain stability.

%\subsection{Illustrative Results}
%\label{subsec:example}
\section{Illustrative Results}
\label{sec:result}

%\subsubsection{Kitaev model}
\subsection{Kitaev model}

We first demonstrate the propagation of HFB wave functions using the Kitaev model, which describes a spinless $p$-wave superconductor. The Hamiltonian is
\begin{equation}
\label{eq:H-Kitaev}
 \hat{H} = -\mu \sum_{i=1}^{L_1} n_i -\sum_{i=1}^{L_1-1}(t c_i^\dagger c_{i+1}+\Delta c_i c_{i+1} + \rm{ h.c.})\,,
\end{equation}
where $\rm{h.c.}$ denotes Hermitian conjugate, $\mu$ is chemical potential, $n_i=c_i^\dagger c_i$ is the number operator, and $L_1$ is the number of sites in the one-dimensional lattice (open boundary condition).
This model can be solved  exactly, since there is no two-body interaction. The ground-state solution has a Majorana energy mode at the boundary \cite{kitaev2001unpaired}.

%However solving 
Solving this %Kitaev 
model by imaginary-time projection is the same as treating one (mean-field) path in the path integral of a many-body Hamiltonian whose HS transformation leads to a one-body Hamiltonian of the form in Eq.~(\ref{eq:H-Kitaev}). It 
 involves all the key elements in generalizing an AFQMC calculation from Slater determinant to 
 HFB states. The only difference with a real QMC calculation is that there is no auxiliary-field to be sampled
 (or put another way, each field can take on a fixed value). The result will therefore be deterministic, with 
 no statistical fluctuation.
%do not need HS transformation to decouple the two-body interaction. QMC results are not statistical which means they have zero variance.
As discussed in Sec.~\ref{sec:QMC}, 
\begin{equation}
|\psi (\tau) \rangle = \exp(-\tau \hat{H}) |\psi_T \rangle
\end{equation}
gives the ground state wave function when $\tau$ is sufficiently large. The ground state energy can be calculated by the mixed estimator
\begin{equation}
 E^M (\tau) =\frac{\langle \psi_T | \hat{H} |\psi(\tau) \rangle}{\langle \psi_T |\psi(\tau) \rangle},
\end{equation}
which involves calculating Green's functions. It can also be calculated by the so-called growth estimator
\begin{equation}
 E^G (\tau) =-\ln [\frac{\langle \psi_T | \exp(-\Delta\tau \hat{H}) |\psi(\tau) \rangle}{\langle \psi_T |\psi(\tau) \rangle} ]\bigg/ \Delta\tau,
\end{equation}
which is usually less costly computationally,  since it only involves calculating overlaps. 
Observables can be computed as full expectation of $|\psi(\tau) \rangle$
\begin{equation}
 \langle \hat{O} \rangle_\tau =  \frac{\langle \psi (\tau) | \hat{O} |\psi(\tau) \rangle}{\langle \psi(\tau) |\psi(\tau) \rangle}.
\end{equation}
%We chose a random wave function as the initial and trial wave function $|\psi_T\rangle$. 

\begin{figure}[htbp]
 \includegraphics[scale=0.40]{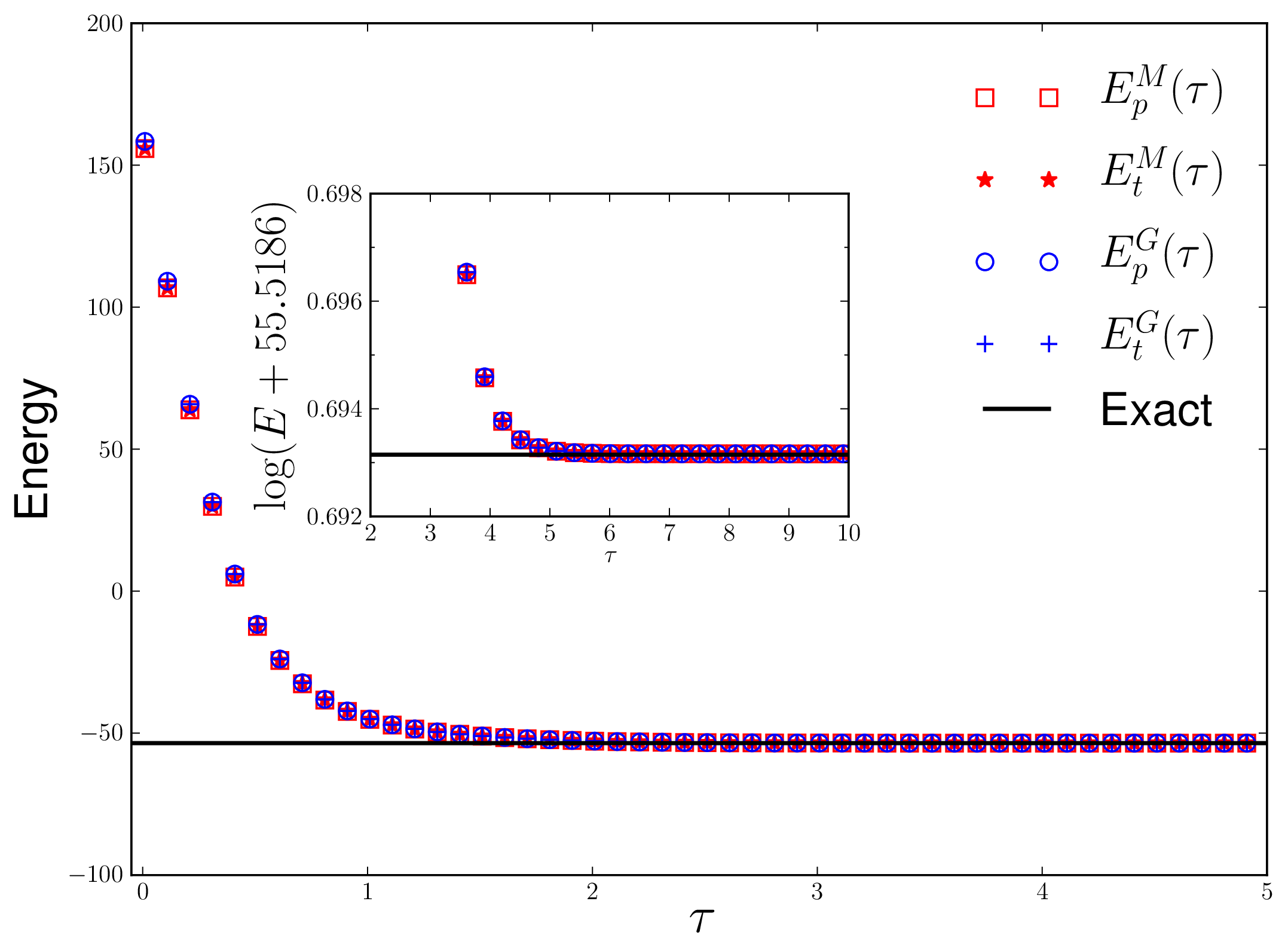}
 \caption{\label{fig:energy}
          (Color online)
          Energy versus imaginary time during projection in the Kitaev model. The lattice size $L_1$ is $100$, 
          and model parameters are $t=1.0$, $\Delta=2.0$, and $\mu=-3.2$. 
          %We choose 
          A time step $\Delta\tau=0.01$ was used. %, and same initial 
          %wave function for product state and Thouless state. 
          Results from propagating product states are numerically the same as those from 
          propagating Thouless states. The mixed estimator and the growth estimator are consistent with each other, and converge to the exact answer for sufficiently large $\tau$.
          The inset shows results for $\tau$ from $2$ to $10$, with log-scale of the energy.
          %when
          %$\tau>3$. Tiny deviation between two estimator come from nonzero time step $\Delta\tau$, and vanish at large imaginary time.
         % \REMARKS{Four symbols are a lit bit crowd. I can add an inset to show $3<\tau<5$, or two insets with each compares two symbols.}
         }
 \end{figure}

\begin{figure}[htbp]
 \includegraphics[scale=0.40]{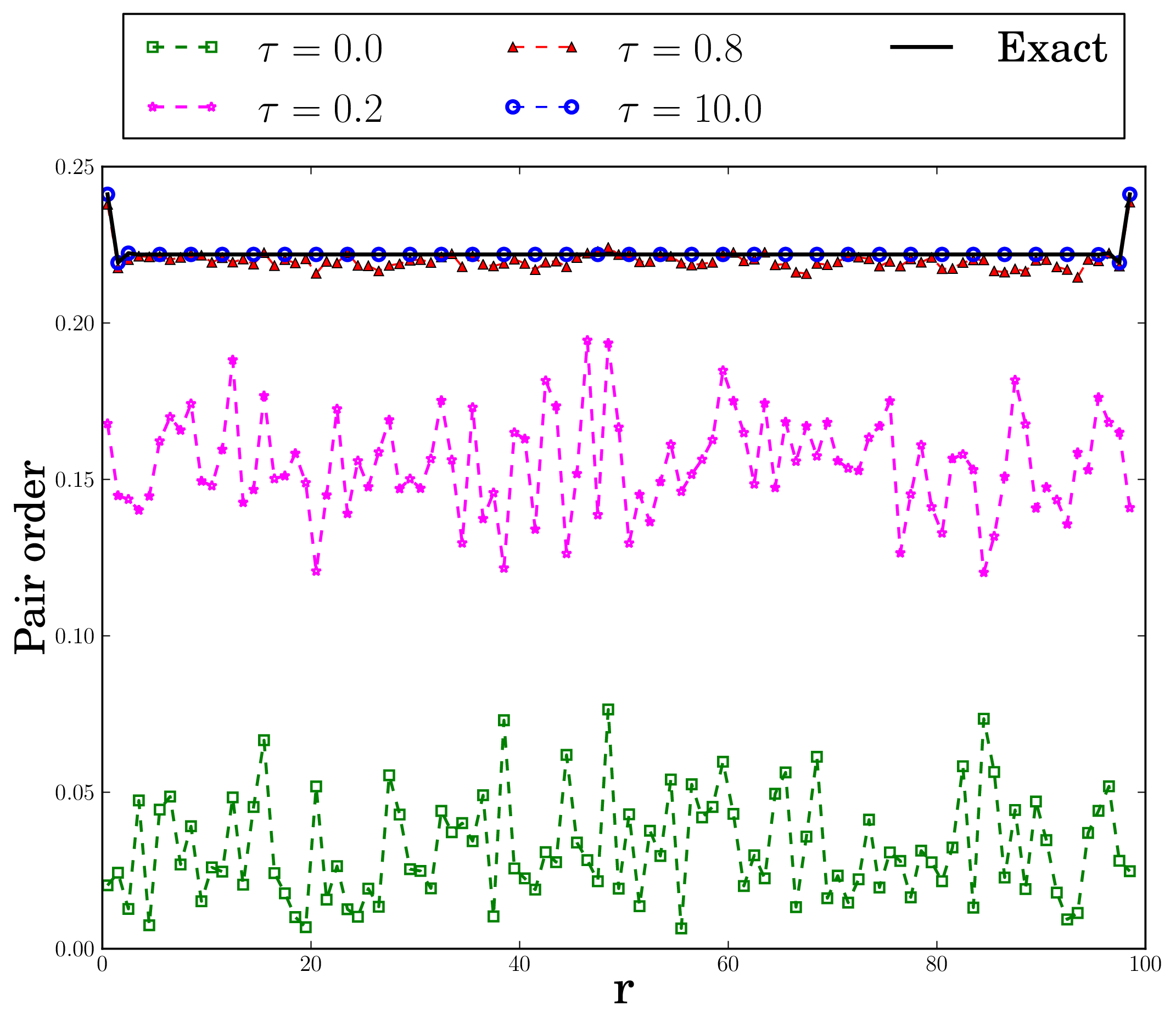}
 \caption{\label{fig:kitaev-pair}
          (Color online) Pairing order $\langle c_r^\dagger c_{r+1}^\dagger \rangle$ vs.~lattice position $r$ in the Kitaev model computed from $|\psi(\tau) \rangle$ at different projection-times $\tau$, with the same parameters in Fig.~\ref{fig:energy}.
          The order parameter converges to the exact solution at the large imaginary time limit.
          For clarity, 
      data in the middle of the lattice are shown at every third value of $r$ for $\tau = 10$.
         }
 \end{figure}

As shown in Fig.~\ref{fig:energy}, the computed energies from product state and Thouless state are numerically equivalent, and both converge to the exact ground-state result at large %when
$\tau$. % is sufficiently large. 
(We use a subscript ``$p$'' or ``$t$'' to indicate results from projection of product state
or Thouless state, respectively. For example, $E_p^M (\tau)$ means the mixed estimator by propagating in the product state form, while $E_t^G (\tau)$ means 
growth estimator by propagating the Thouless state form.)
In these tests, we chose a random wave function as the initial and trial wave function $|\psi_T\rangle$,
which was first set in the product form, and then mapped to the Thouless form. 
The growth estimator has a small deviation with the mixed estimator at small imaginary times, which results from the Trotter error from the nonzero time step size $\Delta\tau$.
The deviation vanishes at large $\tau$ when $|\psi(\tau)\rangle$ becomes the exact ground state.
In Fig.~\ref{fig:kitaev-pair}, we show the computed pairing order at different imaginary times.
The initial value at $\tau=0.0$ is from the random initial wave function. The result is seen to converge to the exact result at the large $\tau$ limit.

%\subsubsection{Hubbard model}
\subsection{Hubbard model}
We next show the propagation of HFB wave functions in an interacting many-fermion system, the  two-dimensional Hubbard model,
\begin{equation}
 \hat{H}_{\rm Hub}=-t\sum_{\langle i,j \rangle \sigma} c_{i\sigma}^{\dagger}c_{j\sigma}+U\sum_{i} n_{i\uparrow}n_{i\downarrow} - \mu \sum_i (n_{i\uparrow} + n_{i\downarrow})\,.
 \label{eq:HubHamil}
\end{equation}
%Here $L$ is the number of lattice sites, 
We will consider periodic lattices with $L_1\times L_2$ sites in the supercell [i.e.,  $N=2(L_1\times L_2)$ in the notation of Eq.~(\ref{eq:general-O-def})]. 
 In Eq.~(\ref{eq:HubHamil}) the sites are labeled by $i$ and $j$,
$c_{i\sigma}^{\dagger}$ and $c_{i\sigma}$ are creation and annihilation operators of an electron
of spin $\sigma$ ($=\uparrow$ or $\downarrow$) on the $i$-th lattice site, $t$ is the nearest-neighbor hopping energy, $U$ is the interaction strength, and $\mu$ is the chemical potential.
We will use $M_\sigma$ %($M_\downarrow$) 
to denote the number of particles with spin $\sigma$.
%up/down.

\begin{figure}[htbp]
 \includegraphics[scale=0.60]{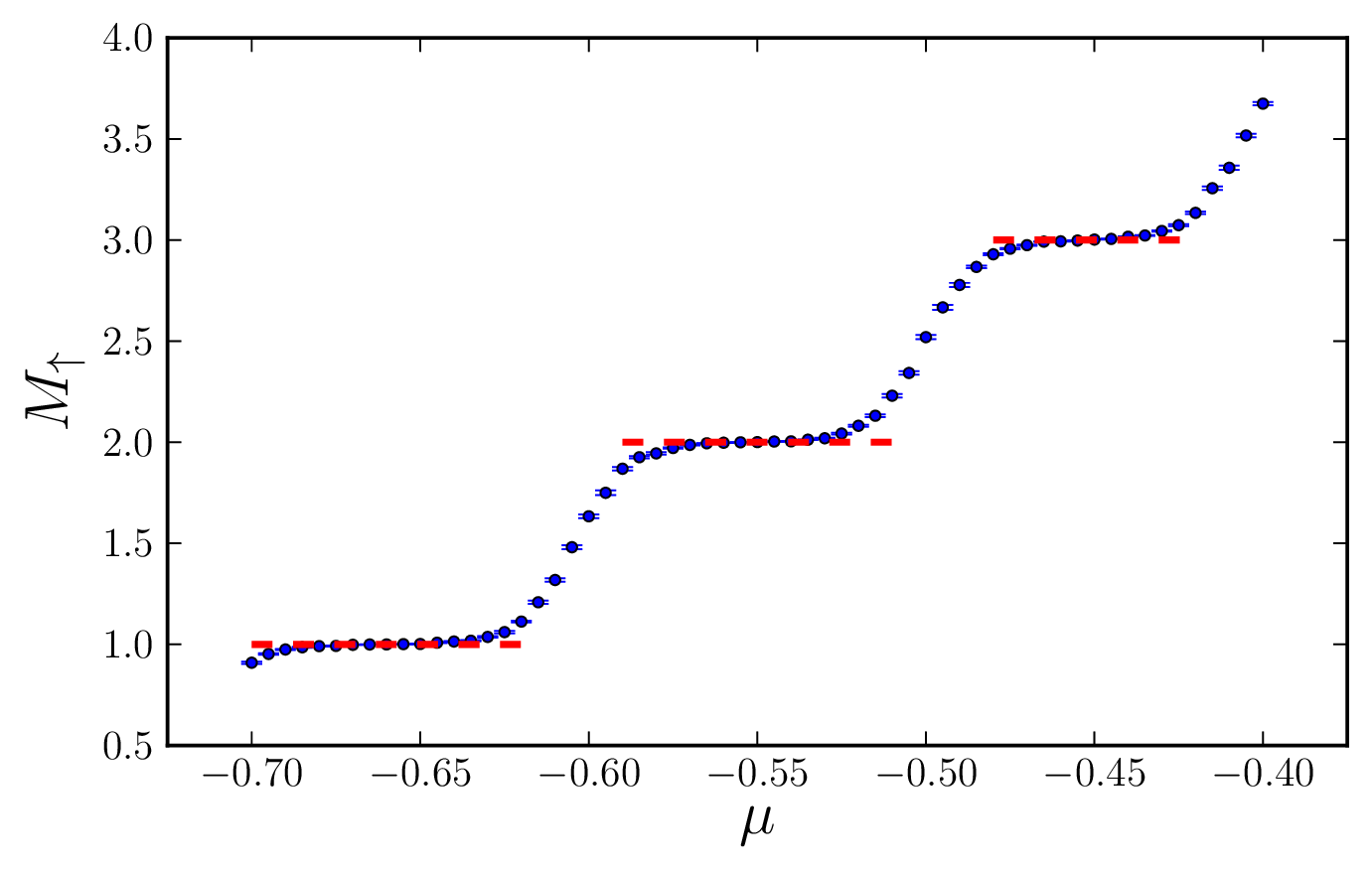}
 \caption{\label{fig:nvsmu}
          (Color online)
          QMC calculations by projecting BCS random walkers.
          Average particle number (for $\uparrow$-electrons) is shown versus chemical potential. The lattice size is $4\times 4$, 
          and model parameters are $t=1.0$, $U=-12.0$. A imaginary-time step of $\Delta\tau=0.01$ was chosen, with projection time $\beta = 64t$. Our BCS initial 
          wave function has $\langle M_\uparrow\rangle= 2.0$. The algorithm converges to different densities
          as $\mu$ is varied and gives accurate results. The plateaus indicate
           integer particle numbers.
                  }
 \end{figure}

\begin{figure}[htbp]
 \includegraphics[scale=0.60]{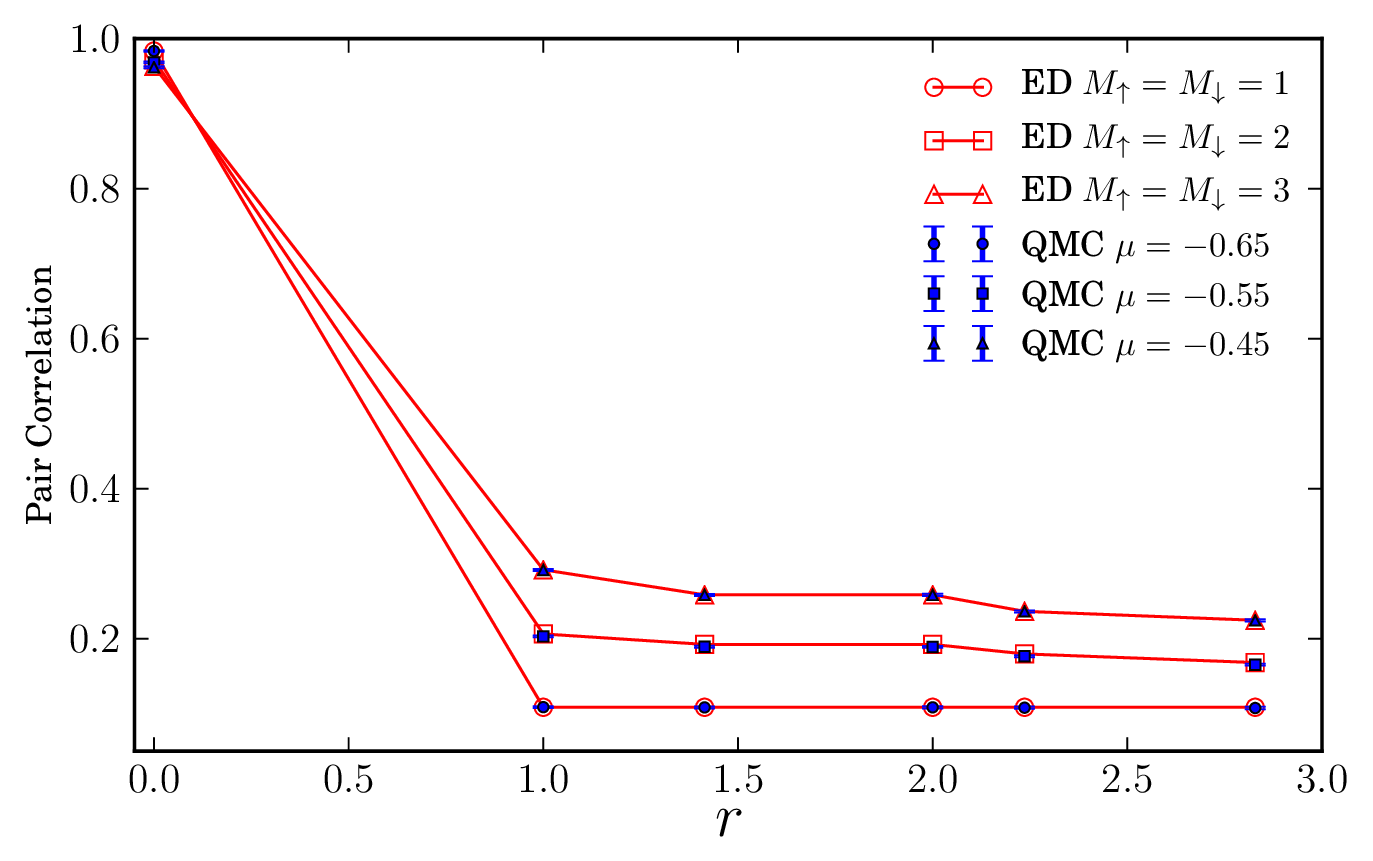}
 \caption{\label{fig:pair-4x4}
          (Color online) Pairing correlation functions computed from QMC and ED. The chemical potential is tuned in QMC to match particle numbers in the
          ED calculations. Same run parameters are used as in Fig.~\ref{fig:nvsmu} 
          and Table~\ref{tab:data}. QMC statistical error bars are smaller than symbol size.
                  }
 \end{figure}

In the attractive Hubbard model ($U<0$), $s$-wave electron pairing is present.  
%the physics can be described by Bardeen–Cooper–Schrieffer (BCS) theory, which is a special case of HFB theory. 
Our initial state will take a Bardeen-Cooper-Schrieffer 
(BCS) wave function, which is a special case of the HFB form.
This wave function is then 
propagated in the AFQMC framework \cite{AFQMC-lecture-notes-2013}, and our trial wave function $|\psi_T\rangle$ 
is also of the BCS form.  
In contrast to 
Slater determinant initial wave functions (such as Hartree-Fock), the number of particles is not conserved in the BCS wave function. The 
chemical potential needs to be tuned to reach the targeted number of particles. In Fig.~\ref{fig:nvsmu}, 
we illustrate the convergence of the QMC propagations of the BCS wave function, and how
the expectation value of the particle number varies as the chemical potential is varied.
(Our calculations are in the $S_z=0$ sector, with $M_\uparrow=M_\downarrow$.) QMC energies are consistent with exact diagonalization (ED) results, as shown in Table~\ref{tab:data}.
We also compute the pairing correlation function \cite{PhysRevA.92.033603} 
\begin{equation}
\label{eqn:pairing-correlation}
 P_{\rm corr} (i) = \langle c_{0\uparrow}^\dagger c_{0\downarrow}^\dagger c_{i\downarrow} c_{i\uparrow}  \rangle\,.
 \end{equation}
 This requires the full estimator which is implemented by back-propagation in the branching randowm walk approach or
 by direct measurement at the middle portions of the path in the path integral formula. Here we used the latter \cite{PhysRevA.92.033603, PhysRevE.93.033303}.
QMC pairing correlation functions are benchmarked against ED results in Fig.~\ref{fig:pair-4x4}
%with respect to 
for different numbers of particles. 

 \begin{table}
\caption{\label{tab:data} Kinetic, interaction, and total 
energies from QMC and ED. 
Three QMC calculations from the middle of the plateaus in Fig.~\ref{fig:nvsmu} are shown,
with $\mu=-0.65$, $-0.55$, and $-0.45$ respectively, which are compared with ED results for fixed 
particle numbers.
The QMC total energy is $\langle \hat H_{\rm Hub} +\mu (M_\uparrow + M_\downarrow)\rangle$.
QMC statistical error bars  are on the last digit and shown in parentheses. }
\begin{tabular}{c|c|c|c|c|c|c}
\hline
\multirow{2}{*}{ $(M_\uparrow, M_\downarrow)$ }& \multicolumn{2}{ c| }{K} & \multicolumn{2}{ c| }{V} & \multicolumn{2}{ c }{E} \\
\cline{2-7}
&ED&QMC&ED&QMC&ED&QMC\\
\hline
$(1,1)$& -2.995  &  -2.997(3) & -10.42  & -10.43(2) & -13.41 & -13.42(2)\\
$(2,2)$& -5.318  &  -5.320(3) & -21.30  & -21.33(2) & -26.62 & -26.65(2)\\
$(3,3)$& -7.162  &  -7.167(4) & -32.46  & -32.42(3) & -39.62 & -39.59(3)\\
\hline
\end{tabular}
\end{table}

The new method affords an advantage in the study of electron pairing correlations, since it
allows one to directly treat a Hamiltonian which contains a pairing field. 
In standard QMC calculations of the Hubbard model (either attractive as in the present case, or repulsive 
in which the $d$-wave pairing correlation is especially of interest), the Hamiltonian does not 
break particle number symmetry, which makes it difficult to directly measure 
%The Hubbard model does not break symmetry of number of particles, which cause problem when we want to measure 
a pairing order parameter, $\langle c_{\uparrow}^\dagger c_{\downarrow}^\dagger \rangle$. 
Typically one instead measures the pairing correlation function in Eq.~(\ref{eqn:pairing-correlation}).

%However, i
If the order parameter is small, $P_{\rm corr} (i)$ will be much smaller since it is related to the square 
of the order parameter at large separation $i$.  
%The pairing order parameters is square root of $P_{corr} (i)$ when $i$ is larger enough. It is a good but not efficient way to estimate 
This makes the task of detecting order especially 
 challenging.   
An alternative way to calculate order parameters is to apply a small pinning field  in the Hamiltonian, and 
detect the order induced by the pinning field \cite{PhysRevLett.112.156403, PhysRevX.3.031010}.
For pairing we could now apply
\begin{equation}
 \hat{H}' = \hat{H}_{\rm Hub} + \sum_i \frac{h_i}{2} (c_{i\uparrow}^\dagger c_{i\downarrow}^\dagger + c_{i\downarrow} c_{i\uparrow})\,,
\end{equation}
where the pairing fields $h_i$ will be non-zero only in a small local region (two neighboring sites in the 
present case).
Using the technique described in this paper, we can solve the above Hamiltonian for the Hubbard model with a pairing pinning field. This was done for up to $16\times 16$ lattices to obtain the $1s$ paring order parameter. 
As illustrated in Fig.~\ref{fig:nvsmu}, %we illustrate the use of the pinning field approach to measure pairing order.
the use of a  pinning field provides a way to measure pairing order with excellent accuracy.
(A more detailed study with finite-size scaling will be required to determine the precise value in the thermodynamic limit.)

 \begin{figure}[htbp]
 \includegraphics[scale=0.60]{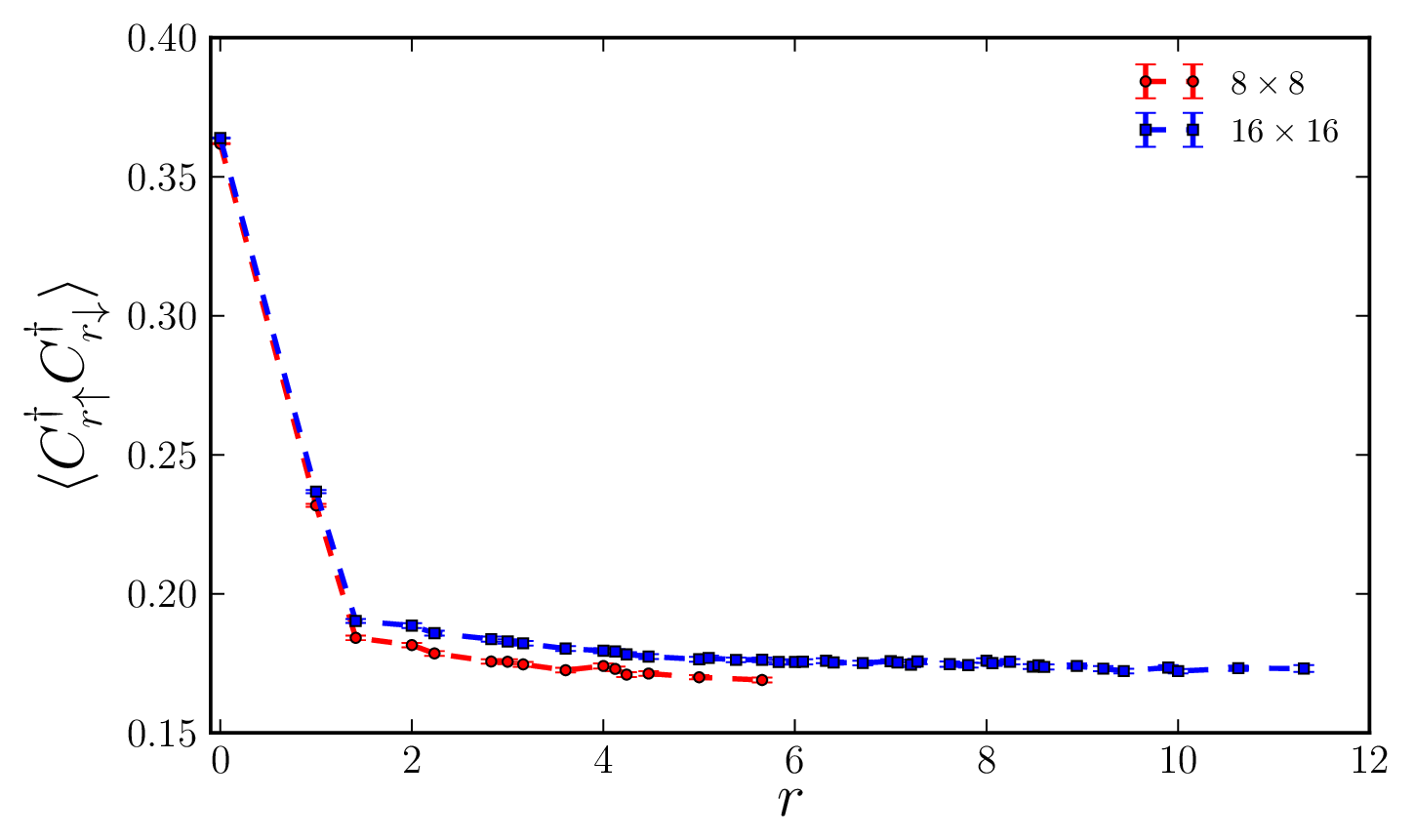}
 \caption{\label{fig:pairing-Hub-pin}
          (Color online) Pairing order versus distance. The lattice sizes are $8 \times 8$ and $16\times 16$, with total number of particles tuned to $10$ and $40$, respectively. The model parameters are $t=1.0$ and $U=-8.0$. 
          We choose a time step $\Delta\tau=0.01$, and projection time $\beta = 64$. Pinning field is put on two neighboring sites (0,0) and (1,0), with $h_i=1$.
          % additional calculation is made to tune chemical potential.
         }
 \end{figure}

\section{Discussion and Summary}
\label{sec:sum}

For clarity, we have separated the  two forms of HFB states, 
the product state and the Thouless state, in the discussion of the technical ingredients. 
The former is more general, while the latter is restricted to fully paired states but gives more compact 
representations. 
Of course they can be mixed and used together as needed, 
both in theory and in numerical implementation. 
A limitation is that we have not implemented or discussed the case of unpaired fermions, 
or when the product in Eq.~(\ref{eq:product-state}) is restricted to a subset of the $N$ quasi-particle operators. We will leave this to a future study.

In Appendix~\ref{appendix:projBCS-det}, we discuss the special example of propagating singlet-pairing 
BCS wave functions, and write out explicit formulas for the ``mixed'' overlap and Green's functions
between a BCS wave function and a Slater determinant. This particular case is useful in the study of 
Fermi gases, for example, where a charge form of the HS decomposition can be 
used to decouple the attractive short-range interaction but a BCS trial wave function greatly 
improves the efficiency \cite{PhysRevA.84.061602}. In this form, the energy can be computed straightforwardly 
with the mixed estimate, but observables require propagating the BCS trial wave function, and keeping
it numerically stable.

%We note that
We have presented the method and formalism in this paper so that they are invariant to whether the Metropolis or the branching random walk method of sampling 
is used, or whether a sign problem is present or not.
The two examples studied in Sec.~\ref{sec:result} are sign-problem-free. When there is a sign or 
phase problem, 
it is straightforward to apply a constraint to control it approximately.  The
constraint is imposed in the branching random walk framework of AFQMC, requiring the calculation of 
the overlap with $|\psi_T\rangle$, and the force bias which is given by the mixed Green's functions. 
Both of these ingredients have been discussed and can be applied 
straightforwardly.

In summary, we have presented  the 
computational ingredients to carry out many-body calculations 
in interacting fermion systems in the presence of 
pairing fields. All aspects required to set up a full QMC calculations in such systems
are described. Components of the formalism presented may also be useful  
in other theoretical and computational contexts and can be adopted. 
We illustrated the method in two situations where propagating a BCS or HFB wave function becomes advantageous 
or even necessary, namely in model Hamiltonians without $U(1)$ symmetry, or  with standard electronic 
Hamiltonians when 
a pairing field term is added to induce superconducting correlations. Related situations include the study of 
Majorana fermions, or in
embedding calculations of standard electronic systems where an impurity is coupled to a bath 
described by a mean-field solution that may have electron
pairing present.

After we have completed a draft of the present work, we became aware of Ref.~\cite{2016arXiv161008022J} which discusses a related approach.

%\section{Summary}
%mention majorana
 
\section{Acknowledgments}

We are grateful to Dr.~S.~Chiesa for many contributions in early stages of this work.
We thank Garnet Chan, Simone Chiesa, Mingpu Qin, Peter Rosenberg, and Bo-xiao Zheng for valuable discussions. This work was supported by 
NSF (Grant no.~DMR-1409510) and the Simons Foundation. Computing was carried out at 
at the Extreme Science and Engineering Discovery Environment (XSEDE), which is supported by National Science Foundation grant number ACI-1053575, and at 
the 
computational facilities at William \& Mary.

\bibliography{phfb}

\appendix 

\section{Additional notations and formulas}
\label{app:addformula}
We first define a matrix representation which will be used throughout the text. Consider a general bilinear operator,
 \begin{equation}
 \label{eqn:generalop}
 \hat{O} = \sum_{ij}^{N} t_{ij} c_i^\dagger c_j + \sum_{i>j}^{N} \Delta_{ij} c_i c_j + \sum_{i>j}^{N} \widetilde{\Delta}_{ij} c_i^\dagger c_j^\dagger + \eta,
\end{equation}
 where $t$, $\Delta$, and $\widetilde{\Delta}$ are corresponding $N\times N$ matrices, and $\eta$ is a constant. Note that $\hat{O}$ can be non-Hermition.
The matrix representation of $\exp(\hat{O})$ is %to be
\begin{equation}
 \exp(\mathbb{O})= \exp    
    \begin{pmatrix}
     t      & \widetilde{\Delta}\\
     \Delta & -t^T
    \end{pmatrix},
\end{equation}
which does not depend on $\eta$,
and we denote its explicit form as
\begin{equation}
\label{eqn:explicit}
 \exp(\mathbb{O})=    
    \begin{pmatrix}
     \mathbb{K}  & \mathbb{M}\\
     \mathbb{K}  & \mathbb{N}
    \end{pmatrix}.
\end{equation}

%\section{Linear Transformation of Quas-particle operator}
%\label{app:lineartrans}
\noindent\emph{\bf Linear Transformation of Quas-particle Operators.}
An arbitrary quas-particle operator $\gamma$ has the form
\begin{equation}
%\gamma = c^\dagger v + cu
\gamma=
    \begin{pmatrix}
     c^\dagger & c
    \end{pmatrix}  
 \begin{pmatrix}
  v \\
  u
 \end{pmatrix},
\end{equation}
with $v=\begin{pmatrix} v_1 & v_2&\dots & v_N \end{pmatrix}^T$ and  $u=\begin{pmatrix} u_1 & u_2 &\dots & u_N \end{pmatrix}^T$.
It can be proven that 
\begin{equation}
\label{eqn:exchange}
 \exp(\hat{O}) \gamma \exp(-\hat{O}) = \gamma^\prime,  % c^\dagger v^{\prime} + cu^{\prime},
\end{equation}
where $\gamma^\prime$ is built from $v^\prime$ and $u^\prime$ with
\begin{equation}
 \begin{pmatrix}
  v^{\prime} \\
  u^{\prime}
 \end{pmatrix}
=
 \exp(\mathbb{O})
 \begin{pmatrix}
  v \\
  u
 \end{pmatrix}.
\end{equation}

To prove the above, we use the expansion
%It is straightforward to expand
\begin{equation}
\label{eqn:linearexpand}
 \exp(\hat{O}) \gamma \exp(-\hat{O}) = \gamma + [\hat{O}, \gamma] + \frac{1}{2!}[\hat{O},[\hat{O}, \gamma]]+\cdots .
\end{equation}
With commutation relations $[\hat{O}, c_j^\dagger] = (c^\dagger t)_j +(c\Delta)_j$ and $[\hat{O}, c_j] = (c^\dagger\widetilde{\Delta})_j+(c (-t^T) )_j $, 
we obtain
\begin{equation}
   [\hat{O}, \gamma] 
   = 
   \begin{pmatrix}
     c^\dagger & c
    \end{pmatrix}
    \begin{pmatrix}
     t      & \widetilde{\Delta}\\
     \Delta & -t^T
    \end{pmatrix}
 \begin{pmatrix}
  v \\
  u
 \end{pmatrix},    
\end{equation}
and 
\begin{equation}
   [\hat{O}, [\hat{O}, \gamma]]
   = 
   \begin{pmatrix}
     c^\dagger & c
    \end{pmatrix}
    \begin{pmatrix}
     t      & \widetilde{\Delta}\\
     \Delta & -t^T
    \end{pmatrix}^2
 \begin{pmatrix}
  v \\
  u
 \end{pmatrix}.
\end{equation}
The right hand side of Eq. (\ref{eqn:linearexpand}) thus gives %is just
\begin{equation}
 \gamma^\prime =    
    \begin{pmatrix}
     c^\dagger & c
    \end{pmatrix}  
    \exp
    \begin{pmatrix}
     t      & \widetilde{\Delta}\\
     \Delta & -t^T
    \end{pmatrix} 
 \begin{pmatrix}
  v \\
  u
 \end{pmatrix}.
\end{equation}

%\section{Expand Exponential Operator}
%\label{app:expand}
\noindent\emph{\bf Expansion of Exponential Operators.}
%K. Hara and S. IWASAKI has talked about expand a unitary exponential operator in paper ... Follow the similar procedure,  we can show the expansion for a general exponential operator.
%
Following Hara and Iwasaki \cite{HARA197961},
% and show the expansion for a general exponential operator.
%For the operator $\hat{O}$ in Eq. (\ref{eqn:generalop}), 
we can expand $\exp(\hat{O})$ to three one-body operators,
%\begin{widetext}
\begin{eqnarray}
\label{eqn:expand}
 \exp(\hat{O}) & = &  \exp(\frac{1}{2} c^\dagger \mathbb{Z} c^{\dagger T}) \exp(c^\dagger \mathbb{X} c^{T}) \exp(\frac{1}{2} c\mathbb{Y} c^{T}) \times  \nonumber \\ 
               &  &  \langle0|\exp(\hat{O}) |0\rangle .
\end{eqnarray}
%\end{widetext}
With the help of matrix representation in Eq.~(\ref{eqn:explicit}), we have
\begin{equation}
\mathbb{Z}  = \mathbb{M} \mathbb{N}^{-1},  
\mathbb{X}  = \ln ( \mathbb{K} ) , 
\mathbb{Y}  = \mathbb{N}^{-1} \mathbb{L}.
\end{equation}
We can also prove
\begin{equation}
 \langle0|\exp(\hat{O}) |0\rangle =\sqrt{\det(\mathbb{N})} \exp[ \frac{1}{2} \Tr (t) +\eta].
\end{equation}

%\section{Compress Exponential Operators}
%\label{app:compress}
\noindent\emph{\bf Compression of Exponential Operators.}
When we have an operator created by multiplying exponentials of one-body operators 
\begin{equation}
\label{eqn:twoexp}
\hat{O}_3= \log [ \exp(\hat{O}_1) \exp(\hat{O}_2) ], 
\end{equation}
$\hat{O}_3$ is still a general one-body operator according to Baker-Campbell-Hausdorff formula.
Its matrix representation is
\begin{equation}
\label{eq:matrix-prod-1body}
 \exp(\mathbb{O}_3) = \exp(\mathbb{O}_1) \exp(\mathbb{O}_2),
\end{equation}
which can be proven by linear transformation relation in Eq. (\ref{eqn:exchange}), 
\begin{eqnarray}
 \gamma^{\prime \prime} & = &\exp(\hat{O}_3) \gamma \exp(-\hat{O}_3) \nonumber  \\ 
                  & = &\exp(\hat{O}_1) [ \exp(\hat{O}_2) \gamma \exp(-\hat{O}_2) ]\exp(-\hat{O}_1),
\end{eqnarray}
where $\gamma^{\prime \prime}$ is built from $v^{\prime \prime}$, $u^{\prime \prime}$ by
\begin{eqnarray}
  \begin{pmatrix}
  v^{\prime \prime} \\
  u^{\prime \prime}
 \end{pmatrix}
&=&
 \exp(\mathbb{O}_1) \exp(\mathbb{O}_2)
 \begin{pmatrix}
  v \\
  u
 \end{pmatrix} \nonumber \\
&=&
 \exp(\mathbb{O}_3)
 \begin{pmatrix}
  v \\
  u
 \end{pmatrix}.
\end{eqnarray}
\COMMENTED{
\REMARKS{I'll come back and work on the rest of this appendix}
With matrix representation, we can determine $\mathbb{t}$, $\mathbb{\Delta}$, and $\widetilde{\mathbb{\Delta}}$ in the operator $\hat{O}_3$, while can not establish $\eta$.  
Since $\eta$ is fixed by Baker-Campbell-Hausdorff formula, we can calculated it from
}
The matrix relations %in Eq.~(\ref{eq:matrix-prod-1body}) 
above define everything up to a proportionality constant. The constant prefactor can be determined from
 \begin{equation}
 \label{eqn:fixeta}
 \langle0| \exp(\hat{O}_3) |0\rangle= \langle0| \exp(\hat{O}_1) \exp(\hat{O}_2) |0\rangle.
\end{equation}
The right-hand side can be calculated by expanding $\exp(\hat{O}_1)$ and $\exp(\hat{O}_2)$  as in 
Eq.~(\ref{eqn:expand}), which leads to overlap of two Thouless state wave functions.

\COMMENTED{
\section{Phase \REMARKS{Phase? Weight? Factor?} in Propagating HFB state}
\label{app:coe}
We talk about calculating phase during propagating an HFB state. 
}

\noindent\emph{\bf Phase of the HFB State After Propagation.}
The phase factor of the product state after propagation is determined by Eq.~(\ref{eqn:alphaproduct}). 
If we have $|\phi \rangle$,  the eigenstate of $ \hat{O} $:
\begin{equation}
 \hat{O} |\phi\rangle = \bar{O} |\phi\rangle,
\end{equation}
it is easy to calculate $\alpha$,
\begin{equation}
\alpha =\exp(\bar{O}) \frac{\langle \phi |  \prod_i \beta_i |0\rangle} {\langle \phi|   \prod_i \beta_i^\prime |0\rangle}\,,
\end{equation}
which only involves two overlaps of HFB wave functions. Alternatively, if 
we choose $|\phi \rangle$ to be the true vacuum, we can apply Eq.~(\ref{eqn:expand}) to expand $\exp(\hat{O})$:
\begin{equation}
\alpha =\langle0|\exp(\hat{O}) |0\rangle \frac{\langle 0 | \exp(\frac{1}{2} c\mathbb{Y} c^{T}) \prod_i \beta_i |0\rangle} {\langle 0|   \prod_i \beta_i^\prime |0\rangle}.
\end{equation}
Exchanging the exponential operator to the right, we obtain
\begin{eqnarray}
\exp(\frac{1}{2} c\mathbb{Y} c^{T})  \prod_i \beta_i |0\rangle & = & \prod_i \beta_i^{\prime \prime}  \exp(\frac{1}{2} c\mathbb{Y} c^{T}) |0\rangle \\
                                                               & = & \prod_i \beta_i^{\prime \prime} |0\rangle,
\end{eqnarray}
so that $\alpha$ can be determined by the overlaps between the true vacumm and HFB states,
\begin{equation}
\alpha =\langle0|\exp(\hat{O}) |0\rangle \frac{\langle 0 | \prod_i \beta_i^{\prime\prime} |0\rangle} {\langle 0|   \prod_i \beta_i^\prime |0\rangle}.
\end{equation}

The phase in Thouless state is determined by Eq.~(\ref{eqn:alphathouless}). When $|\phi \rangle$ is chosen to be the true vacuum, we can expand $\exp(\hat{O})$ as in Eq.~(\ref{eqn:linearexpand}),
\begin{equation}
 \alpha =\langle 0|\exp(\hat{O}) |0 \rangle \langle 0| \exp(\frac{1}{2} c\mathbb{Y} c^{T}) |\psi_t\rangle\,,
\end{equation}
which is given by the %it is just o
verlap of two Thouless state wave functions.

%\section{The special case of a singlet-pairing BCS wave function and a Slater determinant}
\section{The special case of an HFB wave function and a Slater determinant}
\label{appendix:projBCS-det}

A special case of our discussions is an HFB wave function with a Slater determinant (SD). Here the HFB wave function is
\begin{equation}
 |\psi \rangle =\exp( \frac{1}{2} c^\dagger \mathbb{Z} (c^\dagger)^T ) | 0 \rangle,
\end{equation}
and the SD wave function is
\begin{equation}
 |\phi \rangle = \prod_i^{M} \phi_i^\dagger |0\rangle,
\end{equation}
with $\phi_i^\dagger = c^\dagger \phi_i$, and $M$ being the number of fermions.

The overlap between the HFB and SD wave functions is determined by
\begin{equation}
\label{eqn:overlapHFBSD}
 \langle \psi |\phi \rangle = \mathrm{pf} ( \phi^T \mathbb{Z}^* \phi ).
\end{equation}
Setting $Q = \phi^T \mathbb{Z}^\dagger \phi$, we have the Green's functions,
\begin{eqnarray}
\label{eqn:greenHFBSD}
 \rho_{ij}=\frac{\langle \psi |c_i^{\dagger} c_j| \phi \rangle}{\langle \psi | \phi \rangle} &=& ( \mathbb{Z}^\dagger \phi Q^{-1} \phi^T )_{ji} ,\nonumber\\
 \kappa_{ij}=\frac{\langle \psi |c_i c_j| \phi \rangle}{\langle \psi | \phi \rangle} &=& (-\phi Q^{-1} \phi^T)_{ij}  ,\nonumber\\
 \overline{\kappa}_{ij}=\frac{\langle \psi |c_i^{\dagger} c_j^{\dagger}| \phi \rangle}{\langle \psi | \phi \rangle} &=& (-\mathbb{Z}^\dagger + \mathbb{Z}^\dagger \phi Q^{-1} \phi^T \mathbb{Z}^\dagger )_{ij}~.
\end{eqnarray}

\noindent\emph{\bf Projected HFB wave function.} In situations where it is desirable to
%, we would like to 
preserve %U($1$) symmetry, %where 
$U(1)$ symmetry
projected HFB (PHFB) wave function becomes useful. 
For a fixed number of particles $M$,
the PHFB wave function is
\begin{equation}
\label{eqn:PHFB}
% | \psi_\mathrm{PHFB} \rangle = \frac{1}{2^M M!}(c^\dagger Z c^\dagger) ^ M | 0 \rangle.
| \psi_\mathrm{PHFB} \rangle = \frac{1}{2^{M/2} {(M/2)}!}(c^\dagger Z c^\dagger) ^ {M/2} | 0 \rangle.
\end{equation}
The overlap between a PHFB and an SD is the same as Eq.~(\ref{eqn:overlapHFBSD}) and the Green's functions are the same as Eq.~(\ref{eqn:greenHFBSD}).

The propagator for PHFB should not break $U(1)$ symmetry. Let us set $\Delta$ and $\widetilde{\Delta}$ to zero in Eq.~(\ref{eqn:generalop}). The new PHFB wave function
after propagation is
\begin{equation}
 | \psi^\prime_\mathrm{PHFB} \rangle = \exp(\hat{O}) | \psi_\mathrm{PHFB} \rangle,
\end{equation}
and $Z^\prime$ in $| \psi^\prime_\mathrm{PHFB} \rangle$ is
\begin{equation}
 Z^\prime = \exp(t) Z \exp(t^T).
\end{equation}

%\noindent\emph{\bf Spinfull model with singlet pairing.} Let us consider spin-$1/2$ fermions in a basis of size $N_{\rm basis}$. If pairing is only between opposite spins,
\noindent\emph{\bf Spin-$1/2$ model with singlet pairing.} Let us consider spin-$1/2$ fermions in a basis of size $N_{\rm basis}$. If pairing is only between opposite spins,
$\mathbb{Z}$ is specialized to 
\begin{equation}
  \mathbb{Z}=
  \begin{pmatrix}
  0 & \mathbb{Z}_0 \\
  -\mathbb{Z}_0^T & 0
 \end{pmatrix},
\end{equation}
where $\mathbb{Z}_0$ is an $N_{\rm basis} \times N_{\rm basis} $ matrix. If $SU(2)$ symmetry is present, $\mathbb{Z}_0$ is Hermition. The SD wave function is in block diagonal form
\begin{equation}
\phi = 
\begin{pmatrix}
 \phi_\uparrow & 0 \\
 0 & \phi_\downarrow
\end{pmatrix},
\end{equation}
where $\phi_\uparrow$ and $\phi_\downarrow$ are 
%$N_{\rm basis} \times M_\mathrm{h} $ matrices.
$N_{\rm basis} \times M/2 $ matrices.

The overlap between the HFB and SD is reduced to a determinant
\begin{equation}
% \langle \psi | \phi \rangle = (-1)^{ M_\mathrm{h}  (M_\mathrm{h} -1 )/2 }\det (\phi_\downarrow^T \mathbb{Z}_0^\dagger \phi_\uparrow ),
\langle \psi | \phi \rangle = (-1)^{ M/2\,(M/2 -1 )/2 }\det (\phi_\downarrow^T \mathbb{Z}_0^\dagger \phi_\uparrow ),
\end{equation}
which can be calculated efficiently. Note that we can ignore the overall sign here if the number of particles is fixed 
in the calculation. % simulations. %by Lapack library with highly efficiency.
If we set $Q_0 = \phi_\downarrow^T \mathbb{Z}_0^\dagger \phi_\uparrow$, the nonzero Green's functions are
\begin{eqnarray}
 \frac{\langle \psi |c_{i\uparrow}^{\dagger} c_{j\uparrow}| \phi \rangle}{\langle \psi | \phi \rangle} &=& ( \mathbb{Z}_0^* \phi_\downarrow (Q_0^T)^{-1} \phi_\uparrow^T )_{ij} ,\nonumber\\
 \frac{\langle \psi |c_{i\downarrow}^{\dagger} c_{j\downarrow}| \phi \rangle}{\langle \psi | \phi \rangle} &=& ( \mathbb{Z}_0^\dagger \phi_\uparrow Q_0^{-1} \phi_\downarrow^T )_{ij} ,\nonumber\\
 \frac{\langle \psi |c_{i\uparrow} c_{j\downarrow}| \phi \rangle}{\langle \psi | \phi \rangle} &=& (-\phi_\uparrow Q_0^{-1} \phi_\downarrow^T)_{ij}  ,\nonumber\\
 \frac{\langle \psi |c_{i\uparrow}^{\dagger} c_{j\downarrow}^{\dagger}| \phi \rangle}{\langle \psi | \phi \rangle} &=& (\mathbb{Z}_0^* - \mathbb{Z}_0^* \phi_\downarrow (Q_0^T)^{-1} \phi_\uparrow^T \mathbb{Z}_0^* )_{ij}~.
\end{eqnarray}

The corresponding projected HFB wave function is similar to Eq.~(\ref{eqn:PHFB}),
\begin{equation}
% | \psi_\mathrm{PHFB} \rangle = \frac{1}{M!}(c_\uparrow^\dagger Z_0 c_\downarrow^\dagger) ^ M | 0 \rangle,
| \psi_\mathrm{PHFB} \rangle = \frac{1}{(M/2)!}(c_\uparrow^\dagger Z_0 c_\downarrow^\dagger) ^ {M/2} | 0 \rangle,
\end{equation}
where $c_\uparrow^\dagger$ and $c_\downarrow^\dagger$ are the same as $c^\dagger$ except for the spin index.
The general operator in Eq.~(\ref{eqn:generalop}) has the form
\begin{equation}
 t = 
 \begin{pmatrix}
  t_\uparrow & 0 \\
  0 & t_\downarrow
 \end{pmatrix},
\end{equation}
with $\Delta$ and $\widetilde{\Delta}$ equal to zero again. After propagation, the new $Z_0^\prime$ is given by
\begin{equation}
 Z_0^\prime = \exp(t_\uparrow) Z_0 \exp(t_\downarrow^T).
\end{equation}
For a system with $SU(2)$ symmetry, we have $t_\uparrow = t_\downarrow^*$ and $Z_0 = U_0 D_0 U_0^\dagger$, where $U_0$ is a unitary matrix and $D_0$ is a diagonal matrix. The propagation is 
\begin{equation}
 Z_0^\prime = (\exp(t_\uparrow) U_0) D_0 ( \exp(t_\uparrow) U_0 )^\dagger,
\end{equation}
and $Z_0^\prime$ will remain Hermition.
%We can understand the propagation by $U_0^\prime = \exp(t_\uparrow) U_0$, which is similar to 
The propagation can be thought of as $U_0^\prime = \exp(t_\uparrow) U_0$, which is similar to 
propagating an SD wave function. Note that maintaining numerical stability in the propagation will likely 
require additional investigation in these situations.

\COMMENTED{
A special case of our discussions is HFB wave function with Slater determinant. Here we define HFB wave function as 
\begin{equation}
 |\psi \rangle = \prod_i^{N_u} d_{i}^\dagger ~ \exp( \frac{1}{2} c^\dagger Z (c^\dagger)^T ) | 0 \rangle,
\end{equation}
with combination of $N_u$ unpaired states and fully paired Thouless state. Use the same definition as main tex, $d^\dagger$
is a vector, and $d^\dagger = c^\dagger D$.

Our Slater determinant wave function $\phi \rangle$ is just unpaired state,
\begin{equation}
 |\phi \rangle = \prod_i^{N} \phi_i^\dagger |0\rangle,
\end{equation}
here $\phi^\dagger = c^\dagger \phi$.

Overlap between two wave function is $\mathrm{pf} (S) = \langle \psi | \phi \rangle$, with
\begin{equation}
S = 
   \begin{pmatrix}
  \phi^T Z^\dagger \phi&  \phi^T D^*  \\
  -D^\dagger \phi & 0 
 \end{pmatrix},
\end{equation}

Green's function can be calculated by
\begin{equation}
 \frac{\langle \psi | c_i^\dagger c_j |\phi \rangle}{\langle \psi | \phi \rangle } = [
  \begin{pmatrix}
   Z^\dagger & D^* \\
   -D^\dagger & 0
  \end{pmatrix}
  \begin{pmatrix}
   \phi & 0 \\
   0 & 1
  \end{pmatrix}
  S^{-1} 
      \begin{pmatrix}
   \phi^T & 0 \\
   0 & 1
  \end{pmatrix} ]_{ji}.
\end{equation}
}

\end{document}